\documentclass[12pt]{article}

\usepackage{amsfonts}
\usepackage{amsmath}
\usepackage{amssymb}
\usepackage{amsthm}
\usepackage{mathtools}
\usepackage{authblk}
\usepackage{appendix}
\usepackage{bm}
\usepackage{bbm}
\usepackage{booktabs}
\usepackage{dsfont}
\usepackage{gensymb}
\usepackage{natbib}
\usepackage{subcaption}
\usepackage{lineno}
\usepackage{enumitem}
\usepackage[T1]{fontenc}
\usepackage{multirow}
\usepackage{soul}
\usepackage{graphicx,url}
\usepackage[utf8]{inputenc}  
\usepackage[dvipsnames]{xcolor}
\usepackage{tabularx}
\usepackage{graphicx}
\newcommand\smallO{
  \mathchoice
    {{\scriptstyle\mathcal{O}}}
    {{\scriptstyle\mathcal{O}}}
    {{\scriptscriptstyle\mathcal{O}}}
    {\scalebox{.7}{$\scriptscriptstyle\mathcal{O}$}}
  }

\usepackage{algorithm}
\usepackage{algpseudocode}
\algrenewcommand\algorithmicrequire{\textbf{Input:}}  
\algrenewcommand\algorithmicensure{\textbf{Output:}}

\newcommand{\lth}{\ell(\theta)}

\newcommand{\Xk}{X_{[k]}}

\newcommand{\lk}[1]{\ell_{k}(#1)}

\newcommand{\thno}{\theta_0}
\newcommand{\delno}{\delta_0}

\newcommand{\mleth}{\widehat{\theta}}
\newcommand{\mlej}{\widehat{\theta}_k}
\newcommand{\mleb}{\overline{\theta}}
\def\thetabar{\mleb}

\usepackage{cleveref}
\crefname{assumption}{Assumption}{Assumptions}
\crefname{remark}{Remark}{Remarks}
\crefname{proposition}{Proposition}{Propositions}
\crefname{theorem}{Theorem}{Theorems}
\crefname{section}{Section}{Section}
\crefname{lemma}{Lemma}{Lemma}
\crefname{algorithm}{Algorithm}{Algorithms}
\crefname{example}{Example}{Examples}
\crefname{figure}{Figure}{Figure}
\crefname{appendix}{Appendix}{Appendix}
\crefname{table}{Table}{Table}
\crefname{equation}{Equation}{Equation}

\newtheorem{theorem}{Theorem}

\newtheorem{assumption}{Assumption} 
\newtheorem{remark}{Remark} 
\newtheorem{lemma}{Lemma}

\def\var{\mathrm{var}}

\def\bX{X}

\newcommand{\N}{\mathrm{N}} 
\renewcommand{\P}{\mathcal{P}}

\newcommand{\X}{\mathcal{X}}

\def\EE{\mathbb{E}}
\def\PP{\mathbb{P}}
\renewcommand{\O}{\mathcal{O}}
\newcommand{\RR}{\mathbb{R}}

\newcommand{\XX}{\mathbb{X}}

\newcommand{\ZZ}{\mathbb{Z}}
\DeclareMathOperator*{\argmax}{argmax}
\DeclareMathOperator*{\argmin}{argmin}
\DeclareMathOperator*{\bias}{bias}

\def\dd{\mathrm{d}}
\def\thetahat{\widehat{\theta}}

\def\wt{\widetilde}
\def\iid{\stackrel{\text{i.i.d.}}{\sim}}
\def\ind{\stackrel{\text{ind}}{\sim}}

\def\W{\mathrm{W}}
\def\xihat{\widehat{\xi}}
\def\xibar{\overline{\xi}}
\def\Pibar{\overline{\Pi}}
\def\indicator{\mathbb{I}}

\def\tp{\wt{p}}

\graphicspath{ {images/} }

\newcommand{\blind}{1}
\newcommand{\spacing}{1.1}

\begin{document}

\def\spacingset#1{\renewcommand{\baselinestretch}%
{#1}\small\normalsize} \spacingset{1}

\if1\blind
{
  \title{\bf Scalable Bayesian Inference for Time Series via Divide and Conquer}
    \author[1]{Rihui Ou}
    \author[2]{Lachlan Astfalck}
    \author[1,3]{Deborshee Sen}
    \author[1]{David Dunson}
    \affil[1]{Department of Statistical Science, Duke University, Durham, USA}
    \affil[2]{School of Physics, Mathematics and Computation, The University of Western Australia, Crawley, Australia}
    \affil[3]{Google LLC, Bangalore, India}
    
    \setcounter{Maxaffil}{0}
    \renewcommand\Affilfont{\itshape\small}
  \maketitle
} \fi

\if0\blind
{
  \bigskip
  \bigskip
  \bigskip
  \begin{center}
    {\LARGE\bf Scalable Bayesian Inference for Time Series via Divide and Conquer}
\end{center}
  \medskip
} \fi

\bigskip
\begin{abstract}
Bayesian computation often scales poorly with increasing data size, motivating developments such as divide-and-conquer approaches for scalable inference. These methods partition the data into subsets, perform parallel inference on each subset, and aggregate the results into a single posterior. Appealing theoretical properties and practical performance have been demonstrated for independent data; however, methods for dependent data remain challenging. Existing methods rely on ad hoc approximations with limited theoretical guarantees that lead to potentially poor accuracy in practice. Here, we focus on time-series data and propose a simple, scalable divide-and-conquer method for dependent time series, with theoretically rigorous accuracy guarantees. Simulation studies and real-data examples are used to empirically verify the effectiveness of our approach.
\end{abstract}

\noindent%
\vfill

\newpage
\spacingset{\spacing}

\section{Introduction} \label{sec:intro}

Recent advances in sensor technology, computing power, and data storage have made the collection of massive datasets increasingly routine. This, in turn, has necessitated the development of scalable algorithms for statistical inference. Although approximate methods that tout scalability, such as variational inference \citep{beal2003variational,blei2017variational} and sequential Monte Carlo \citep{del2006sequential}, have grown in popularity, Markov chain Monte Carlo (MCMC) remains the default for most practitioners. Unfortunately, MCMC scales at least linearly with data size, and for dependent data the cost can be substantially higher; for instance, computation often grows cubically in the case of exact likelihood computation for Gaussian processes. This renders MCMC impractical for the data sizes typical of modern scientific and industrial applications. Divide-and-conquer strategies offer a promising alternative, whereby the data are partitioned, parallel inference is performed on each subset using MCMC, and the resulting subset posteriors are aggregated into a single posterior distribution. Although such methods have been well studied for independent data, scalable inference for dependent observations remains challenging. We propose a novel divide-and-conquer approach for long time series under weak dependence assumptions. The method is theoretically justified with rigorous approximation error bounds and is straightforward to implement in practice.

The main alternative to divide-and-conquer for scalable MCMC is subsampling. Rather than partitioning the data, these methods achieve scalability by modifying the MCMC transition kernel itself \citep{ma2015complete,quiroz2018speeding,nemeth2020stochastic}. At each iteration of the MCMC sampler, the likelihood (and its gradients, where applicable) is estimated using a random subset of the data, rather than the full dataset. Such methods have been developed for both Langevin dynamics \citep{welling2011bayesian} and Hamiltonian dynamics \citep{chen2014stochastic}. Initial developments focused on independent and identically distributed observations, with extensions since proposed for hidden Markov models \citep[HMMs;][]{ma2017stochastic,aicher2019stochastic_1}, general stationary time series \citep{salomone2020spectral,villani2024spectral}, and, more recently, nonlinear state-space models \citep{aicher2025stochastic}. In general, subsampling methods introduce approximations to the transition kernel whose effect on the stationary distribution can be difficult to characterize, and theoretical guarantees typically require verifying challenging conditions on a case-by-case basis. A promising recent approach uses non-reversible continuous-time samplers, such as piecewise-deterministic Markov processes, that preserve the exact posterior for independent data \citep{bouchard2018bouncy,bierkens2019zig}, but rely on gradient upper bounds that limit practical applicability. Despite these advances, results in \cite{johndrow2020no} highlight fundamental limitations of subsampling methods in the context of large-scale inference.

Rather than modifying the MCMC transition kernel, divide-and-conquer methods achieve scalability through parallel computational architecture. The core contribution is to allow parallelization of MCMC computations across subsets of the data, such that distributed computing resources may be exploited. Within divide-and-conquer methods, the main distinction lies in the strategy by which the subset posteriors are aggregated. Here, we focus on methods based on the Wasserstein barycenter \citep[][]{li2017simple,srivastava2018scalable}, and provide mathematical details and comparisons with related literature in Section~\ref{sec:comparison}. Historically, most divide-and-conquer algorithms have been developed for independent data, with limited work extending them to dependent settings. \citet{guhaniyogi2017divide} develop a related approach for spatial data, but rely on restrictive assumptions, including Gaussianity. Further, \cite{wang2023divide} propose a divide-and-conquer approach for finite state-space HMMs. We are motivated by the need for scalable inference in long time series under serial dependence, while relaxing some of the assumptions in prior work. For such settings, conventional MCMC and sequential Monte Carlo algorithms are often computationally infeasible. There is a rich literature on alternative strategies, ranging from variational approximations \citep{johnson2014stochastic,foti2014stochastic} to assumed density filtering \citep{lauritzen1992propagation}. In general, these approaches lack theoretical guarantees on posterior accuracy and can substantially underestimate uncertainty in practice. When distributed computing resources are available, divide-and-conquer methods offer a compelling and theoretically grounded alternative.

In this article, we develop a simple, broadly applicable, and theoretically supported class of Wasserstein barycenter-based divide-and-conquer methods for massive time series. We consider general time-series models and do not require hidden Markov structure, or Gaussian assumptions. We refer to our methodology as divide-and-conquer for Bayesian time series (DC-BATS). We consider stationary, ergodic, short-memory time series, and demonstrate asymptotic convergence of the aggregated posterior from DC-BATS to the true posterior. Under the additional assumption that the maximum-likelihood estimator is unbiased, we show convergence at the optimal rate of $T^{-1/2}$. Further, we supply asymptotic guarantees on the bias and variance of the DC-BATS posterior.

The rest of the article is organized as follows. Section~\ref{sec.dc-bats} introduces our proposed method, DC-BATS. Section~\ref{sec.theory} is devoted to a theoretical analysis of the method. In particular, we show that DC-BATS returns asymptotically exact estimates of projections of the posterior distribution. Next, Section~\ref{sec.synthetic.data} demonstrates the proposed method via simulation studies on a class of time-series models with flexible dependence properties. We apply the proposed method to a real-data example of Los Angeles particulate matter in Section~\ref{sec.pm.dataset}. Finally, Section~\ref{sec.discussion} concludes the article. Additional model assumptions are given in the Appendix, and all proofs are contained in the Supplementary Material. 

\section{Divide-and-conquer for time series}
\label{sec.dc-bats}

\subsection{A generic time-series model}

We consider a stochastic process indexed by time. Let $\ZZ$ denote the set of integers, and for each $t \in \ZZ$, let $X_t$ be a random variable taking values in a measurable space $(\XX, \X)$, where $\XX \subseteq \mathbb{R}^k$ is the state space and $\X$ is its Borel $\sigma$-algebra. The collection $\{X_t : t\in \ZZ\}$ forms a stochastic process. For any $t_1 \leq t_2$, we use the notation $X_{t_1:t_2} \coloneq (X_{t_1}, \dots, X_{t_2})$; in particular, $X_{1:T} \coloneq (X_1, \dots, X_T)$ denotes the full observed time series of length $T$. We assume that the data are generated from a parametric model $p_\theta$, with $\theta = (\theta_1,\dots,\theta_d) \in \Theta \subseteq \RR^d$. Specifically, the model defines a marginal distribution $p_\theta(X_1)$ and a sequence of conditional distributions $p_\theta(X_t \mid X_{1:(t-1)})$ for all $t \in \{2, \dots, T\}$. We assume all such distributions admit densities with respect to a reference measure on $(\XX, \X)$, corresponding to the Lebesgue measure. The posterior distribution will likewise be assumed to admit a density with respect to Lebesgue measure on \(\mathbb{R}^d\).

The log-likelihood of any temporal sequence of observations $X_{1:T}$ can be expressed as
\begin{equation} \label{eq.full.likelihood}
\lth
= 
\log p_\theta(X_1) + \sum_{t=2}^T \log p_\theta(X_t \mid X_{1:(t-1)}),
\end{equation}
which includes the special case of independent observations when $p_\theta(X_t \mid X_{1:(t-1)}) = p_\theta(X_t)$. In this article, we focus on dependent data settings, encompassing classical ARMA models (and related members of the time-series acronym family), conditionally heteroskedastic processes, and latent-state models such as HMMs. Bayesian inference proceeds by specifying a prior distribution $\Pi_0(\dd \theta)$ on $\theta$ and computing the posterior distribution 
\begin{equation} \label{eq.full.posterior}
\Pi_T (\dd \theta \mid X_{1:T}) 
\propto 
p_\theta(X_1) \left \{ \prod_{t=2}^T p_\theta(X_t \mid X_{1:(t-1)}) \right \} \Pi_0(\dd \theta),
\end{equation}
which we refer to as the \emph{full posterior}, as it is conditioned on the entire dataset. Samples from the full posterior in \cref{eq.full.posterior} may be obtained using standard MCMC algorithms, including Metropolis-Hastings \citep{metropolis1953equation,hastings1970monte}, the Metropolis-adjusted Langevin algorithm (MALA; \citealp{roberts1996exponential}) and Hamiltonian Monte Carlo (HMC; \citealp{duane1987hybrid}). However, since the log-likelihood in \cref{eq.full.likelihood} must be evaluated at every iteration, computation becomes increasingly expensive as $T$ grows. Moreover, for very large $T$, it may not be feasible to store or manipulate the entire dataset on a single machine, making full-data MCMC impractical. To address this, we propose an embarrassingly parallel divide-and-conquer strategy for scalable Bayesian inference in time series.

\subsection{Divide-and-conquer algorithm} \label{sec.dc.algo}

In general, divide-and-conquer algorithms proceed by partitioning the $T$ observations into $K$ disjoint subsets of sizes $m_1, \dots, m_K$, such that $\sum_{k=1}^K m_k = T$. For notational simplicity, and without loss of generality, we assume equal partition sizes, $m_1 = \cdots = m_K = T/K \coloneq m$. For independent observations, the partition may be arbitrary; however, for time-series data, we must partition the observations sequentially to preserve the temporal structure. Accordingly, we define the $k$th \emph{subsequence} as $\Xk \coloneq X_{(k-1)m+1:km}$, so that the full dataset is given by the ordered collection $X_{1:T} = (X_{[1]}, \dots, X_{[K]}).$

For each subsequence $\Xk$, we define the \emph{pseudo-likelihood} $\tp_\theta(\Xk)$, an approximation to the full-data likelihood that ignores dependence on earlier observations, effectively treating \(X_{[k]}\) as though it begins at time one. For the first subsequence, this approximation is exact such that $\tp_\theta(X_{[1]}) = p_\theta(X_{[1]}) = p_\theta(X_1) \prod_{t=2}^m p_\theta(X_t \mid X_{1:(t-1)})$. For $k \in \{2, \dots, K\}$, the pseudo-likelihood is defined as
\begin{align} \label{eq.subset.prob}
\tp_\theta(\Xk) 
& = 
p_\theta(X_{(k-1)m+1}) \prod_{t=(k-1)m+2}^{km} p_\theta \left( X_t \mid X_{{(k-1)m+1}:t-1} \right).
\end{align}
We now define the \emph{subsequence posteriors} $\Pi(\dd \theta \mid \Xk)$ for each subsequence $\Xk$ as 
\begin{equation} \label{eq.subset.posterior}
\Pi (\dd \theta \mid \Xk)
=
\frac{\tp_\theta (\Xk)^{\gamma_k} \Pi_0(\dd\theta)}{\int_\Theta \tp_\theta (\Xk)^{\gamma_k} \Pi_0(\dd \theta)},
\end{equation}
where the indices $\gamma_1,\dots,\gamma_K > 0$ control the effective number of observations in the pseudo-likelihoods. For instance, if $K=1$ and $\gamma_1 = 1$, \cref{eq.subset.posterior} recovers the full posterior in \cref{eq.full.posterior}. More generally, for $K$ subsequences, setting $\gamma_1 = \cdots = \gamma_K = 1$ results in \cref{eq.subset.posterior} giving the standard posteriors for each subsequence, but these are based on fewer observations and thus are not on the same scale as the full posterior. In the theory that follows, we assume the process $\{X_t: t \in \mathbb{Z}\}$ is stationary,  so the distribution $\tp_\theta (\Xk)$ is the same for all $k$. Heuristically, this suggests that each subsequence carries approximately the same amount of information, and the full dataset $X_{1:T}$ carries $K$ times as much information as any individual subsequence. This leads to a natural choice of $\gamma_1 = \cdots =\gamma_K = T/m = K$. We support this heuristic argument with theoretical results in \cref{sec.theory} and numerical experiments in \cref{sec.synthetic.data}. 

\subsection{Aggregation with the Wasserstein barycenter}

Divide-and-conquer methods primarily differ in how they aggregate the subsequence posteriors into a single approximation of the full posterior in \cref{eq.full.posterior}. In this work, we adopt the Wasserstein barycenter of the set of subsequence posteriors as the aggregation rule. Let $\P_2(\Theta)$ denote the set of all probability measures on $\Theta \subseteq \RR^d$ with finite second moments. The Wasserstein-2 distance between probability measures $\mu, \nu \in \P_2(\Theta)$ is defined as 
\begin{equation*}
\W_2(\mu,\nu) 
=
\left \{ \inf_{\lambda \in \Lambda(\mu,\nu)} \int_{\Theta \times \Theta} \| \theta_1 - \theta_2 \|^2 \lambda(\dd \theta_1 \, \dd \theta_2) \right \}^{1/2},
\end{equation*}
where $\Lambda(\mu,\nu)$ denotes the set of all probability measures on $\Theta \times \Theta$ with marginals $\mu$ and $\nu$, respectively, and $\|\cdot\|$ denotes the Euclidean norm on \(\mathbb{R}^d\). Convergence in $\W_2$ distance on $\P_2(\Theta)$ is equivalent to weak convergence plus convergence of the second moment \citep[Lemma 8.3]{bickel1981some}. The Wasserstein barycenter of the subsequence posteriors is defined as
\begin{equation} \label{eq.wass.posterior}
\Pibar(\dd \theta \mid \bX_{1:T})
=
\underset{\mu \in \P_2(\Theta)}{\argmin} 
\sum_{k=1}^{K} \W_2^2(\mu, \Pi_{[k]}),
\end{equation}
where we use the shorthand $\Pi_{[k]} \coloneq \Pi (\dd \theta \mid \Xk)$. We assume that $\Pibar(\dd \theta \mid \bX_{1:T})$ admits a density $\overline{\pi}(\theta \mid \bX_{1:T})$ with respect to the Lebesgue measure, so that, $\Pibar(\dd \theta \mid \bX_{1:T}) = \overline{\pi}(\theta \mid \bX_{1:T}) \, \dd \theta$.

Although other distances between probability measures may be used for aggregation, the Wasserstein-2 distance is particularly appealing, as it describes the geometric center of the subsequence posteriors \citep{agueh2011barycenters,srivastava2015wasp}. In particular, \citet{agueh2011barycenters} established existence and uniqueness of the Wasserstein barycenter under general conditions, and \citet{srivastava2015wasp} proved strong consistency. In addition, \citet{szabo2019asymptotic} demonstrated that Wasserstein barycenter-based posteriors can exhibit favorable asymptotic properties. The approach was applied in the independent data setting by \citet{li2017simple}, who combined subset posteriors using a Wasserstein barycenter under a factorized likelihood. However, their framework does not directly extend to time-series data, where such likelihood factorizations are unavailable. In this work, we show that defining pseudo-likelihoods for each subsequence and aggregating the resulting posteriors via \cref{eq.wass.posterior} yields provably accurate approximations to the full posterior, even in the presence of serial dependence.

Exact computation of the Wasserstein barycenter is an NP-hard problem, and efficient approximation methods remain an active area of research \citep[e.g.,][]{cuturi2014fast, dvurechenskii2018decentralize}. However, for one-dimensional functionals of the parameter \(\theta\), the Wasserstein barycenter admits a simple analytic form based on averaged quantiles. Let $a \in \RR^d$, $b \in \RR$, and define the scalar projection $\xi = a^\top \theta + b$. Let $\Pi(\xi \mid X_{[k]})$ and $\bar{\Pi}(\xi \mid X_{1:T})$ denote the marginal cumulative distribution functions (CDFs) of $\xi$ induced by the posterior distributions $\Pi(\dd \theta \mid \Xk)$ and $\Pibar(\dd \theta \mid \bX_{1:T})$, respectively. For any $u \in (0,1)$, the corresponding quantile functions are defined by
\begin{align*}
\Pi^{-1}(u \mid X_{[k]}) &:= \inf \{ \xi \in \mathbb{R} : u \leq \Pi(\xi \mid X_{[k]}) \}, \quad \text{and}\\
\bar{\Pi}^{-1}(u \mid X_{1:T}) &:= \inf \{ \xi \in \mathbb{R} : u \leq \bar{\Pi}(\xi \mid X_{1:T}) \}.
\end{align*}
For such linear functionals $\xi$, the quantile function of the Wasserstein barycenter is
\[
\bar{\Pi}^{-1}(u \mid X_{1:T}) = \frac{1}{K} \sum_{k=1}^K \Pi^{-1}(u \mid X_{[k]}).
\]
This identity enables fast computation of credible intervals for one-dimensional summaries, or projections, of \(\bar{\Pi}(\dd \theta \mid X_{1:T})\) using only quantiles from the subsequence posteriors. Given samples from the subsequence posterior distributions, these quantiles can be efficiently estimated via standard Monte Carlo techniques. A summary of the full divide-and-conquer procedure is provided in Algorithm~1.

\spacingset{1.1}
\begin{algorithm}[t!] \label{alg.dc-bats}
\caption{DC-BATS: Divide-and-Conquer for Bayesian Time Series}
\begin{algorithmic}[1]

\Require Time series $X_{1:T}$; prior $\Pi_0(\dd \theta)$; subsequence length $m$; number of subsequences $K = T / m$, indices $\gamma_1, \dots, \gamma_K$ with default value $K$.
\Ensure Samples from divide-and-conquer posterior $\bar{\Pi}(\dd \theta \mid X_{1:T})$

\State \textbf{Partition data:} Divide the time series sequentially into $K$ disjoint subsequences:
\[
X_{[k]} \gets X_{(k-1)m + 1 : km}, \quad \text{for } k = 1, \dots, K
\]

\State \textbf{Construct pseudo-likelihoods:} For each subsequence $X_{[k]}$, define
\[
\tilde{p}_\theta(X_{[k]}) \gets p_\theta(X_{(k-1)m + 1}) \prod_{t = (k-1)m + 2}^{km} p_\theta\left(X_t \mid X_{(k-1)m + 1 : t-1}\right)
\]

\State \textbf{Form subsequence posteriors:}
\[
\Pi(\dd \theta \mid X_{[k]}) \gets \frac{\tilde{p}_\theta(X_{[k]})^{\gamma_k} \Pi_0(\dd \theta)}{\int_\Theta \tilde{p}_\theta(X_{[k]})^{\gamma_k} \Pi_0(\dd \theta)}
\]

\For{\(k = 1\) to \(K\)}
  \State Draw samples $\{\theta^{(k)}_j\}_{j=1}^N$ from $\Pi(\dd \theta \mid X_{[k]})$ using MCMC or another sampler
\EndFor

\State \textbf{Aggregate:} Compute the Wasserstein barycenter of the subsequence posteriors:
\[
\bar{\Pi}(\dd \theta \mid X_{1:T}) \gets \argmin_{\mu \in \mathcal{P}_2(\Theta)} \sum_{k=1}^K \mathrm{W}_2^2(\mu, \Pi_{[k]})
\]

\State \textbf{Optional (1D functional):} For scalar functionals \(\xi = a^\top \theta + b\), compute
\[
\bar{\Pi}^{-1}(u \mid X_{1:T}) \gets \frac{1}{K} \sum_{k=1}^K \Pi^{-1}(u \mid X_{[k]}), \quad u \in (0,1)
\]
\State \Comment{Provides quantiles for \(\bar{\Pi}(\xi \mid X_{1:T})\) without full barycenter computation}

\end{algorithmic}
\end{algorithm}
\spacingset{\spacing}
\begin{remark}
If the data exhibit finite-order dependence of order $r$, such that $p_\theta(X_t \mid X_{1:t-1}) = p_\theta(X_t \mid X_{t-r:t-1})$, then the full log-likelihood admits an exact decomposition across overlapping blocks. In this case, one could construct a proper likelihood-based divide-and-conquer method using buffered subsequences, and combine the resulting posteriors through the Wasserstein average, similar to \citet{li2017simple}. For a fixed and finite $r$, the computational complexity of such an approach 
scales the same to that of our proposed method. However, the assumption of finite-order dependence corresponds to a restricted model class, subsumed by the more general dependence structures that we study. We conjecture that similar theoretical results to those in \cref{sec.theory.main_results} may also be obtained in this special case.
\end{remark}

\subsection{Comparison with relevant literature} \label{sec:comparison}

A historical challenge in divide-and-conquer for Bayesian inference is how to combine subset posteriors into a coherent approximation to the full posterior. Early work leveraged the product form of the joint posterior to combine subset posteriors via kernel smoothing \citep{neiswanger2013asymptotically}, multiscale histograms \citep{wang2015parallelizing}, or Weierstrass approximations \citep{wang2013parallelizing}. \cite{scott2016bayes} propose a consensus Monte Carlo algorithm, which averages draws from each subset posterior and can be justified under approximate normality. \cite{minsker2014scalable} instead proposed aggregation via the geometric median of subset posteriors. More recently, Wasserstein barycenters have emerged as an attractive aggregation strategy with strong theoretical properties. In particular, \citet{li2017simple} showed that Wasserstein-based posteriors can be computed efficiently for independent data and often provide better uncertainty quantification than averaging-based methods. \cite{srivastava2015wasp} established strong consistency of Wasserstein posterior barycenters, and \cite{szabo2019asymptotic} proved that they achieve optimal posterior contraction rates and yield credible sets with correct frequentist coverage in certain settings. These approaches are all \emph{embarrassingly parallel}, in that they involve no communication between the subset posteriors beyond a single unification step.

Extensions of divide-and-conquer methodology to dependent data remain relatively limited. \citet{guhaniyogi2017divide} proposed a method for spatial data using Gaussian process priors, but their approach is limited to the Gaussian case and only provides error bounds on posterior means (in turn, implying error rates for the $L_2$ risk). In contrast, our method applies to general time-series models and provides Wasserstein convergence guarantees for the full posterior, which in turn imply rates for both posterior bias and variance. \citet{wang2023divide} introduced a divide-and-conquer algorithm for finite-state hidden Markov models using a double-parallel Monte Carlo strategy \citep{xue2019double}, whereas we adopt Wasserstein barycenter aggregation and allow for generic time-series dependence and models for which likelihoods need not be tractable via forward-backward filtering. Alternative strategies based on subsampling frequency-domain approximations have also been developed. For example, \citet{salomone2020spectral} and \citet{villani2024spectral} propose subsampling MCMC algorithms based on the Whittle likelihood \citep{whittle1951hypothesis}. However, such approaches are limited to second-order stationary models and are less flexible in handling missing data, latent-state structures, or irregularly observed time series. Finally, \citet{dai2023bayesian,dai2019monte} develop non-embarrassingly parallel algorithms for independent and identically distributed data that avoid aggregation bias but require global communication steps, making them less suitable for large-scale or distributed implementations.

In contrast to this literature, our proposed method is fully embarrassingly parallel, generalizes to dependent data without assuming finite-state, finite-order dependence or Gaussian structure, and offers both practical scalability and theoretical guarantees. Our algorithm raises each subsequence likelihood to an appropriate power to match the information content of the full data, then combines the resulting posteriors using the Wasserstein barycenter. This results in a provably accurate posterior approximation, even when exact MCMC samples are available only for each subsequence. We now state these theoretical results.

\section{Theoretical guarantees}
\label{sec.theory}

\subsection{Notation}

We assume a true parameter value $\thno \in \Theta \subseteq \RR^d$, in the sense of Proposition~6.7 of \cite{ghosal2017fundamentals}, and that the process $\{X_t : t \in \mathbb{Z}\}$ is generated by the probability measure $\PP_{\thno}$ induced by $\thno$. Expectations under this measure are denoted by $\EE_{\thno}$. Let $\lk{\theta} = \log \tp_\theta(\Xk)$ denote the \emph{pseudo log-likelihood} of the $k$th subsequence, where $\tp_\theta(\Xk)$ is defined as in \cref{eq.subset.prob}. We use the shorthand $\nabla_\theta$ for the gradient operator $\partial/(\partial \theta)$, so that $\nabla_\theta \ell(\theta)$ and $\nabla^2_\theta \ell(\theta)$ denote the gradient and Hessian, respectively. Define  $\mlej = \argmax_{\theta \in \Theta} \lk\theta$ as the maximum likelihood estimator of the $k$th subsequence, and let $\mleb = \sum_{k=1}^K \mlej/K$ denote the average across the $K$ subsequence MLEs. The MLE based on the full dataset is denoted $\mleth=\argmax_{\theta \in \Theta} \ell_T(\theta)$ where \(\ell_T(\theta)\) is the full data log-likelihood from \cref{eq.full.likelihood}. We assume that all MLEs are uniquely defined.

Throughout, we measure vector and matrix magnitudes using standard norms: the Euclidean norm $\|v\|=( \sum_{i=1}^d v^2_i )^{1/2}$ for vectors $v\in \RR^d$, and the Frobenius norm $\|V\|=( \sum_{i=1}^d\sum_{j=1}^d V^2_{ij} )^{1/2}$ for matrices $V\in \RR^{d\times d}$. For a real-valued random variable $Z$, we define its $L^p$-norm as $\|\cdot\|_p = \{\EE_{\thno}(|\cdot|^p) \}^{1/p}$. For a sequence of real-valued random variables $\{Z_n\}_{n\geq 1}$, each  measurable with respect to the $\sigma$-algebra generated by $\{X_t : t \in \mathbb{Z}\}$, and a deterministic sequence $\{a_n\}_{n \geq 1}$, we write $Z_n = \mathcal{O}_{\thno}(a_n)$ if for any $\epsilon>0$, there exists $s>0$ and $N \in \mathbb{N}$ such that $\PP_{\thno}(|Z_n / a_n|>s) < \epsilon$ for all $n>N$. Similarly, we write $Z_n = \smallO_{\thno}(a_n)$ if $ \PP_{\thno}(|Z_{n}/a_n| > \epsilon) \rightarrow 0$ as $n \rightarrow \infty$, for any $\epsilon>0$. Finally, we write $\sigma(\{X_t : t \in \mathbb{Z}\})$ for the smallest $\sigma$-algebra with respect to which all $X_t$ are measurable.

\subsection{Main assumptions}
\label{sec.notations.assumptions}

We now provide the main assumptions used in our theoretical proofs. For conciseness, additional technical conditions, similar to those in \cite{li2017simple}, are deferred to \cref{sec.additional_assumptions}. The assumptions here concern stationarity and ergodicity, mixing time, and decay of the conditional score function.
\begin{assumption}[Stationarity and ergodicity] \label{as.staerg}
The process $\{X_t: t \in \mathbb{Z}\}$ is strictly stationary and ergodic.
\end{assumption}
\begin{assumption}[Mixing time] \label{as.mixing}
The $\alpha$-mixing coefficient of $\{X_t: t \in \mathbb{Z}\}$,
\begin{equation*}
\alpha(n)
=
\sup _{A \in \sigma(\dots,X_{-1},X_0), \, B \in \sigma(X_n, X_{n+1},\dots)}|\PP_{\thno}(A) \PP_{\thno}(B)-\PP_{\thno}(A \cap B)|,
\end{equation*}
satisfies $\sum_{j=1}^\infty \alpha(j)^{\delta /(2+\delta)}<\infty$ for some $\delta>0$. 
\end{assumption}
\begin{assumption}[Decay of the conditional score function] \label{as.mixscorefun}
There exist constants $\rho_0 \in (0,1)$, $C_0 > 0$, and a sufficiently large integer $N$ such that
\begin{align*}
\EE_{\thno} \Big[ \|\nabla_\theta \log p_{\thno}(X_1 \mid X_{-i:0})-\nabla_\theta\log p_{\thno}(X_1 \mid X_{-j:0})\|_1 \Big]
& \leq
C_0 \, \rho_0^{N},
\end{align*}
for all $i,j>N$.
\end{assumption}
\cref{as.mixing} excludes processes with long-range dependence. The assumption is mild and satisfied by a wide class of weakly dependent processes. In particular, it holds for geometrically ergodic processes: if there exists a $0 < \rho < 1$ such that $\alpha(n) < \rho^n$ for all sufficiently large $n$, then $\sum_{n=1}^\infty \alpha(n)^{\delta /(2+\delta)} \leq \sum_{n=1}^\infty \{\rho^{\delta /(2+\delta)}\}^n < \infty$. \cref{as.mixscorefun} states that, in expectation, the difference between conditional score functions given increasingly distant pasts vanishes at a geometric rate. Intuitively, this condition enforces a notion of memory in the model, ensuring that distant past observations have vanishing influence on the current data's contribution to the likelihood. This assumption is trivially satisfied by finite-order Markov processes. For an $n$-order Markov process, the conditional distribution $p_{\thno}(X_1 \mid X_{-i:0})$ depends only on $X_{-(n-1):0}$ for all $i \geq n$, and so for any $i,j > n$,
\[
\EE_{\thno} \big[ \|\nabla_\theta \log p_{\thno}(X_1 \mid X_{-i:0}) - \nabla_\theta \log p_{\thno}(X_1 \mid X_{-j:0}) \|_1 \big] = 0.
\]
The condition also holds for more complex dependent processes. For example, in finite state-space HMMs, Lemma 6 of \cite{bickel1998asymptotic} guarantees the existence of a limiting score $\eta_1$ such that $\EE_{\thno} \big[ \|\nabla_\theta \log p_{\thno}(X_1 \mid X_{-i:0}) - \eta_1 \|_1 \big] \leq C_0 \rho_0^i$, for some constants $C_0 > 0$ and $\rho_0 \in (0,1)$. This implies for $i,j > N$ and via the triangle inequality, 
\begin{align*}
\EE_{\thno} \big[ \|\nabla_\theta \log p_{\thno}(X_1 \mid X_{-i:0}) - \nabla_\theta \log p_{\thno}(X_1 \mid X_{-j:0}) \|_1 \big]
& \leq
2 C_0 \rho_0^N,
\end{align*}
which satisfies \cref{as.mixscorefun}. Importantly, the assumption is not limited to Markov or HMM models, but applies to any process for which the conditional score function stabilizes as the conditioning history grows. The same geometric-decay argument applies to finite order moving average models with independent innovations, as well as to geometrically ergodic GARCH and stochastic-volatility models under standard regularity conditions. However, the assumption excludes processes with truly infinite memory or long-range dependence such as ARFIMA models.

\subsection{Main results} \label{sec.theory.main_results}

We now present the main theoretical results of the paper. Our results consider the asymptotic regime where the total time-series length $T = Km \rightarrow \infty$ with the number of subsequences $K$ growing and the subsequence length $m$ also tending to infinity, albeit at a slower rate. We show that the error introduced by combining the subsequence posteriors using DC-BATS vanishes asymptotically. The proofs are available in the Supplementary Material, and extend the results of \cite{li2017simple} to dependent time series and under broader assumptions. We first establish \cref{lem.meandif}, a novel result that is instrumental in proving our later results.  Recall that $\mlej$ is the MLE of the $k$th subsequence, $\thetabar = \sum_{k=1}^K \mlej/K$ is the average MLE across the $K$ subsequences, and $\mleth$ is the MLE from the full dataset.
\begin{lemma}\label{lem.meandif}
Under Assumptions 1--3 and Assumptions 4--9 in \cref{sec.additional_assumptions}, the average of the subsequence MLEs, $\mleb$, satisfies $\|\mleb - \mleth\| = \smallO_{\thno}(m^{-1/2})$. Further, if we additionally assume that each subsequence MLE $\widehat{\theta}_k$ is unbiased for $\theta$, i.e., $\EE_{\thno}(\widehat{\theta}_k) = \theta_0$, and if $m = \mathcal{O}(T^{1/2})$, then $\|\mleb - \mleth\| = \smallO_{\thno}(T^{-1/2})$. In this case, the averaged estimator converges to the full data MLE at the standard parametric rate.
\end{lemma}

The first result of \cref{lem.meandif} holds under general regularity conditions and does not require unbiasedness. The second, sharper rate relies on both an unbiasedness assumption and a specific growth regime for the subsequence size $m$. \cref{thm.main}, which leverages \cref{lem.meandif}, is the first main theoretical result of this paper.
\begin{theorem}[Error due to combining subsequence posteriors] \label{thm.main}
Suppose Assumptions 1--3 and Assumptions 4--9 in \cref{sec.additional_assumptions} hold. Let $\xi=a^\top \theta+b$ for fixed $a \in \RR^{d}$ and $b \in \RR$; denote $\xibar = a^\top \thetabar+b$ and $\xihat=a^\top \thetahat+b$; and define $I_{\xi}(\thno) = [a^\top \{I^{-1}(\thno)\} a]^{-1}$ as the Fisher information  for the functional $\xi$ at $\thno$. The following results hold as $m \to \infty$ and $T \to \infty$.

\begin{enumerate}
\item[(a)] Denote by $\Phi(\cdot ; \mu, \Sigma)$ the normal distribution with mean $\mu$ and variance $\Sigma$. Then,
\begin{align*}
T^{1/2} \W_2 \Big(\Pibar(\dd \xi \mid X_{1:T}) \; , \;  \Phi(\dd \xi ; \xibar, T^{-1} I_{\xi}(\thno)^{-1}) \Big)
& \to 
0 ,
\\
T^{1/2} \W_2 \Big( \Pi(\dd \xi \mid X_{1:T})\; , \;  \Phi(\dd \xi ; \xihat, T^{-1} I_{\xi}(\thno)^{-1}) \Big) 
& \to 
0,
\\
m^{1/2} \W_2 \Big(\Pibar(\dd \xi \mid X_{1:T}) \; , \; \Pi(\dd \xi \mid X_{1:T}) \Big) 
& \to 
0.
\end{align*}
\item[(b)]
If the additional conditions in Lemma~\ref{lem.meandif} hold, that is, $\EE_{\thno} (\thetahat_k) = \thno$ and
$m = \O(T^{1/2})$,
then
\[T^{1/2} \W_2\Big(\Pibar(\dd \xi \mid X_{1:T}) \; , \; \Pi(\dd \xi \mid X_{1:T})\Big) \to 0.\]
\end{enumerate}
All convergences are with respect to $\PP_{\thno}$-probability.
\end{theorem}

For \cref{thm.main}a to hold, it suffices that $m \to \infty$ at a much slower rate than $T$. The more interesting result is \cref{thm.main}b, which establishes that the Wasserstein distance between the aggregated posterior $\Pibar_T$ and the full posterior $\Pi_T$ achieves the optimal convergence rate of $T^{-1/2}$, provided the subsequence MLEs $\thetahat_k$ are unbiased and $m = \O(T^{1/2})$. This requirement is not restrictive in practice: divide-and-conquer algorithms typically partition the data into a fixed number $K$ of subsets (often in the tens or hundreds), while $T$ is much larger. Empirically, we observe that DC-BATS performs well even when both $T$ and $m$ are moderate, suggesting that the asymptotics are effective at relatively small sample sizes. Finally, \cref{thm.main} also enables accuracy guarantees for posterior moments, which we formalize in the following \cref{thm.moments}.

\begin{theorem}[Guarantees on first and second moments] 
\label{thm.moments}
Let $\xi = a^\top \theta + b$ for fixed $a \in \RR^{d}$ and $b \in \RR$, and define $\xi_0 = a^\top \theta_0 + b$. Define the posterior bias of a distribution $\Pi(\dd \xi \mid X_{1:T})$, $\bias\left[\Pi(\dd \xi \mid X_{1:T})\right] = \int \xi \, \Pi(\dd \xi \mid X_{1:T}) - \xi_0$. Under Assumptions 1--3 and Assumptions 4--9 in \cref{sec.additional_assumptions}, the following conditions hold.
\begin{enumerate}
\item[(a)] The posterior bias satisfies
\begin{align*}
\bias\left[\Pibar(\dd \xi \mid X_{1:T})\right] &= \xibar - \xi_0 + \smallO_{\thno}(T^{-1/2}) \\ 
\bias\left[\Pi(\dd \xi \mid X_{1:T})\right] &= \xihat - \xi_0 + \smallO_{\thno}(T^{-1/2}).
\end{align*}
\item[(b)] The posterior variances satisfy
\begin{align*}
\var\left[\Pibar(\dd \xi \mid X_{1:T})\right] &= T^{-1} I_{\xi}^{-1}(\thno) + \smallO_{\thno}(T^{-1}), \\  
\var\left[\Pi(\dd \xi \mid X_{1:T})\right] &= T^{-1} I_{\xi}^{-1}(\thno) + \smallO_{\thno}(T^{-1}).
\end{align*}
\end{enumerate}
\end{theorem}

\cref{thm.moments} quantifies the asymptotic bias and variance of both the aggregated and full posteriors. While bias is traditionally a frequentist notion, we adopt a version adapted to the Bayesian setting, following the definition in \cite{li2017simple}. Unlike the classical (fixed) bias, the quantity considered here is random, as it depends on the data through the posterior. The difference in bias between the aggregated and full posteriors is governed by $\xibar - \xihat$, up to an asymptotically negligible term of order $\smallO_{\thno}(T^{-1/2})$. More generally, \cref{lem.meandif} shows that this difference is $\smallO_{\thno}(m^{-1/2})$ under minimal assumptions, and improves to $\smallO_{\thno}(T^{-1/2})$ under mild regularity conditions. As for posterior variance, both posteriors agree on the dominating term $T^{-1} I^{-1}_{\xi}(\thno)$, and differ only by an asymptotically negligible remainder of order $\smallO_{\thno}(T^{-1})$.

\begin{remark}
Our theoretical results concern the exact Wasserstein barycenter, whereas in practice one would typically compute the barycenter of Monte Carlo approximations to the subsequence posteriors. As discussed in  \cite{li2017simple}, the additional error due to Monte Carlo sampling can be accounted for and is typically negligible relative to the asymptotic errors described above.
\end{remark}

\section{Synthetic data experiments} \label{sec.synthetic.data}

We evaluate DC-BATS through a series of simulation studies. As illustrations, we report results for two representative cases: (i) an autoregressive model, exemplifying a classical short-memory process, and (ii) an autoregressive tempered fractionally integrated moving average (ARTFIMA) model, which exhibits semi-long-range dependence. Previous work has shown that stochastic gradient MCMC algorithms systematically underestimate posterior variances, leading to poor uncertainty quantification (see, e.g., Figure~2 of \citealp{nemeth2020stochastic}). Moreover, they face fundamental trade-offs between scalability and accuracy \citep{johndrow2020no}. For these reasons, our comparisons focus on DC-BATS versus full-data MCMC. Additional experiments are provided in the Supplementary Material, including for nonstationary settings and models with low to moderate $T$ and $m$.

\subsection{Linear regression with autoregressive errors} 
\label{sec.app.ar.error}

\begin{figure}[ht]
\centering
\includegraphics[width=\textwidth]{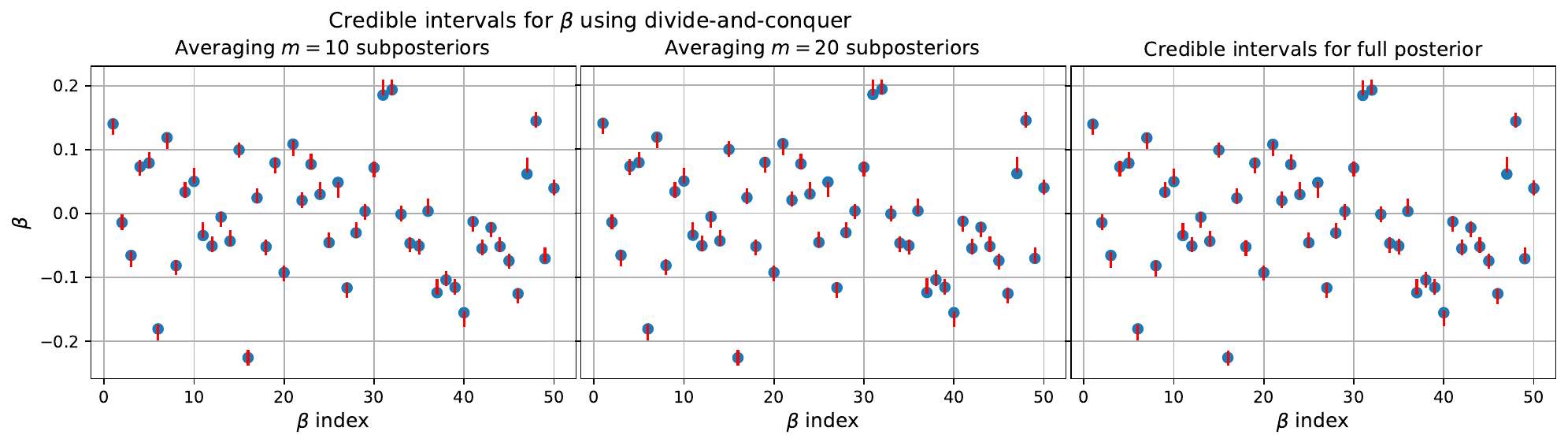} 
\caption{Ninety-five percent credible intervals for the regression coefficients $\beta$ in the linear regression model with autoregressive errors \eqref{eq.model.1}. Results are shown for DC-BATS with $K \in \{10,20\}$ subsequences and for the full-data MCMC.}
\label{fig.CI_ar(2)_errors}
\end{figure}

We first consider a linear regression model with autoregressive errors,
\begin{align} \label{eq.model.1}
\begin{aligned}
X_t 
&= \alpha + \beta^\top Z_t + \varepsilon_t, 
\\ 
\varepsilon_t
&= \varphi_1 \varepsilon_{t-1} + \varphi_2 \varepsilon_{t-2} + \xi_t,
\quad \text{with } \xi_t \iid \N(0,\sigma^2),
\end{aligned}
\end{align}
where $X_t, \alpha, \varepsilon_t \in \RR$, $\beta, Z_t \in \RR^p$, and the initial conditions satisfy $\varepsilon_0, \varepsilon_{-1} \ind \N(0,\sigma^2)$. Here, $X_{1:T}$ denotes the outcome observations and $Z_{1:T}$ the covariates. Initially, we set $(\varphi_1, \varphi_2) = (0.4, -0.6)$, choose $p=50$, and generate $T=10^5$ observations from this model. For inference, independent $\N(0,10^2)$ priors are placed on $\alpha, \varphi_1, \varphi_2$, and on each component of $\beta$. We further specify an inverse-gamma $\mathrm{IG}(3,10)$ prior on $\sigma^2$. We consider $K \in \{10,20\}$ subsequences, drawing $10^4$ samples from each subsequence posterior as well as from the full posterior using the no-U-turn sampler (NUTS; \citealp{hoffman2014no}) as implemented in Stan \citep{carpenter2017stan}. The first half of the samples are discarded as burn-in. Figure~\ref{fig.CI_ar(2)_errors} plots 95\% credible intervals for $\beta$. The intervals produced by DC-BATS are virtually indistinguishable from those obtained via full-data MCMC. Frequentist coverage of the credible intervals for $\beta$ is 94\% for $K=10$, 92\% for $K=20$, and 94\% under the full posterior.

\spacingset{1.1}
\spacingset{\spacing}

Next, to represent different degrees of dependence, we consider four parameter settings for $(\varphi_1, \varphi_2)$: (a) i.i.d., $(\varphi_1, \varphi_2) = (0.0, 0.0)$; (b) Case~I, $(0.3, 0.1)$; (c) Case~II, $(0.8, 0.0)$; and (d) Case~III, $(0.4, 0.4)$. We set $T=10^5$ and $p=10$, and for each $K \in \{5,10,20\}$ generate 50 datasets. This yields 500 credible intervals for each combination of $K$ and $(\varphi_1, \varphi_2)$. \cref{tab:nested_table} reports the empirical coverage of 95\% credible intervals. Across all scenarios, DC-BATS attains frequentist coverage comparable to that of full-data MCMC. Here we study an AR-only process, but since moving average components with independent innovations also satisfy our assumptions, the same conclusions will extend to the more general class of ARMA processes.

\spacingset{1.1}
\begin{table}[ht]
\centering
\begin{tabularx}{\textwidth}{|c|XX|XX|XX|XX|}
\hline
\multirow{2}{*}{} & \multicolumn{2}{c|}{i.i.d.} & \multicolumn{2}{c|}{Case I} & \multicolumn{2}{c|}{Case II} & \multicolumn{2}{c|}{Case III} \\
\cline{2-9}
& DC & Full & DC & Full & DC & Full & DC & Full \\
\hline
\( K=5 \) & 94 & 95 & 93 & 96 & 94 & 96 & 96 & 94  \\
\hline
\( K=10 \) & 92 & 95 & 95 & 96 & 94 & 96 & 94 & 94 \\
\hline
\( K=20 \) & 92 & 95 & 94 & 96 & 96 & 96 & 94 & 94\\
\hline
\end{tabularx}
\caption{Frequentist coverage (\%) of 95\% credible intervals for DC-BATS and full-data MCMC across four scenarios: i.i.d., Case~I, Case~II, and Case~III. Coverage is reported for $K \in \{5,10,20\}$ subsequences.}

\label{tab:nested_table}
\end{table}
\spacingset{\spacing}

\subsection{ARTFIMA time series and practical limitations} 

We now assess the performance of DC-BATS for data generated from an ARTFIMA process of \cite{meerschaert2014tempered}. This process introduces a tempering parameter $\lambda$ that interpolates between long-memory ARFIMA behaviour ($\lambda \to 0$) and short-memory ARMA behaviour ($\lambda \to \infty$), and is often described as exhibiting semi-long memory \citep{goodwin2024dynamic}. An \text{ARTFIMA}$(p,d,\lambda,q)$ process is defined by
\[
\Phi_p(B)\,\Delta^{d, \lambda} X_t \;=\; \Psi_q(B)\,\varepsilon_t, 
\qquad \varepsilon_t \stackrel{\text{i.i.d.}}{\sim} \mathcal{N}(0,\sigma^2),
\]
where $B$ is the backward-shift operator, $\Phi_p(B) = 1 - \sum_{i = 1}^p \phi_i B^i$ and $\Psi_q(B) = 1 + \sum_{i = 1}^q \psi_i B^i$ are the standard autoregressive and moving-average lag-polynomials, and $\Delta^{d, \lambda}$ is the tempered fractional differencing operator, defined as
\begin{equation*}
    \Delta^{d, \lambda} = (1 - \exp\{-\lambda\}B)^d = \sum_{k=0}^{\infty} {d \choose k } \, \exp\{-\lambda k\}\, B^{k}
\end{equation*}
for $\lambda \geq 0$. The process is stationary when $-0.5 < d < 0.5$ and the roots of $\Phi_p(B)$ and $\Psi_q(B)$ lie outside the unit circle. In what follows, we study the \text{ARTFIMA}$(1,d,\lambda,0)$ specification with $\sigma = 1$.

We investigate the effect of different levels of tempering $\lambda \in \{0.005,\,0.1\}$, fractional integration $d\in\{0.1,\,0.3\}$, and autoregressive parameter $\phi\in\{0.1,\,0.9,\,0.99\}$ on the ability of DC-BATS to recover the full posterior. For each parameterization, we generate $100$ independent series of length $T = 10^{5}$, treat $\lambda$ as known, and estimate $d$ and $\phi$. The full posterior is compared against the DC-BATS posteriors aggregated across $K\in\{5,\,10,\,20\}$ partitions. As a performance metric, we compute the \emph{normalized Wasserstein distance},
\begin{equation*}
    \mathrm{norm \, W}_1\Big(\Pibar(\dd \xi \mid X_{1:T}) \; , \; \Pi(\dd \xi \mid X_{1:T})\Big) = \frac{\mathrm{ W}_1\Big(\Pibar(\dd \xi \mid X_{1:T}) \; , \; \Pi(\dd \xi \mid X_{1:T})\Big)}{\mathrm{ W}_1\Big(\Pi(\dd \xi \mid X_{1:T}) \; , \; \mathrm{m}_\xi\Big)},
\end{equation*}
where $\mathrm{m}_\xi$ denotes the posterior median of $\xi$. This corresponds to the Wasserstein-1 distance (Earth mover's distance) between the full and DC-BATS posteriors, rescaled by the mean absolute deviation of the full posterior around its median. The resulting statistic is dimensionless and scale-invariant, and allows comparisons across parameter settings. An interpretation of a score of $\mathrm{norm \, W}_1 = c$ is that on average, $\Pibar(\dd \xi \mid X_{1:T})$ differs from $\Pi(\dd \xi \mid X_{1:T})$ by $c$ mean absolute deviations. Results are reported in Table~\ref{table.artfima} for $\xi = \phi$.

There are two clear trends in these results: (1) as $K$ increases, and (2) as the persistence $\phi$ increases, the performance of DC-BATS worsens. This is entirely expected and highlights the practical feasibility of our method. Our main results in Section~\ref{sec.theory} are asymptotic, and require $m = T/K \to \infty$. Write the full log-likelihood as
\begin{equation*}
  \log p_\theta(X_{1:T}) = \sum_{k=1}^K \log p_\theta(X_{[k]}) + \sum_{k = 1}^{K-1} \underbrace{\log \frac{p_\theta(X_{[k+1]} \mid X_{1:km})}{p_\theta(X_{[k+1]})}}_{\coloneqq \mathcal{D}_k},
\end{equation*}
where the $\mathcal{D}_k$ represent the cross-subsequence dependence. Under $\alpha$-mixing (Assumption 2), $\mathcal{D}_k = \mathcal{O}_{\theta_0}(1)$ and so the per-sample error in studying independent subsequences can be quantified as
\begin{equation*}
  \frac{1}{T}\left[ \sum_{k = 1}^{K-1} \log \frac{p_\theta(X_{[k+1]} \mid X_{1:km})}{p_\theta(X_{[k+1]})} \right] = \mathcal{O}_{\theta_0}(K/T) = \mathcal{O}_{\theta_0}(1/m) \to 0.
\end{equation*}
For finite samples, it follows that larger $K$ induces larger error. Further, for highly persistent processes (e.g. the $\phi = 0.99$ case), the leading constants in these error terms are inflated. In practice, this suggests using the smallest feasible number of partitions $K$, with the understanding that more persistent time series demand larger sample sizes to ensure accurate approximation.

\spacingset{1.1}
\begin{table}[ht]
\centering
\begin{tabular}{ccc|ccc}
\toprule
 $\phi$&   $d$&$\lambda$&K = 5& K = 10&K = 20\\
\midrule
\multirow{4}{*}{0.1} & \multirow{2}{*}{0.1} & 0.005 & 0.14 & 0.23 & 0.32 \\
\cmidrule(lr){3-3}
& & 0.1 & 0.22 & 0.31 & 0.51 \\
\cmidrule(lr){2-3}
& \multirow{2}{*}{0.3} & 0.005 & 0.17 & 0.30 & 0.63\\
\cmidrule(lr){3-3}
& & 0.1 & 0.19 & 0.36 & 0.64 \\
\midrule 
\multirow{4}{*}{0.9}& \multirow{2}{*}{0.1} & 0.005 & 0.33 & 0.75 & 1.9 \\
\cmidrule(lr){3-3}
& & 0.1 & 0.32 & 0.69 & 1.5 \\
\cmidrule(lr){2-3}
& \multirow{2}{*}{0.3} & 0.005 & 0.37 & 0.82 & 1.6 \\
\cmidrule(lr){3-3}
& & 0.1 & 0.30 & 0.65 & 1.3 \\
\midrule
\multirow{4}{*}{0.99} & \multirow{2}{*}{0.1} & 0.005 & 0.78 & 1.44 & 2.8 \\
\cmidrule(lr){3-3}
& & 0.1 & 0.74 & 1.5 & 2.9 \\
\cmidrule(lr){2-3}
& \multirow{2}{*}{0.3} & 0.005 & 0.70 & 1.5 & 2.8 \\
\cmidrule(lr){3-3}
& & 0.1 & 0.60 & 1.4 & 2.7 \\
\bottomrule
\end{tabular}
\caption{Normalized Wasserstein-1 distance between the full posterior and the DC-BATS posteriors, averaged across 100 simulations. Data are generated from \text{ARTFIMA}$(1,d,\lambda,0)$ processes with tempering $\lambda \in \{0.005,0.1\}$, fractional integration $d \in \{0.1,0.3\}$, and autoregressive coefficient $\phi \in \{0.1,0.9,0.99\}$. Results are reported for numbers of partitions $K \in \{5,10,20\}$.}
\label{table.artfima}
\end{table}
\spacingset{\spacing}

\section{Application to Los Angeles particulate matter data} 
\label{sec.pm.dataset}

\begin{figure}[ht]
\centering
\includegraphics[width=0.95\textwidth]{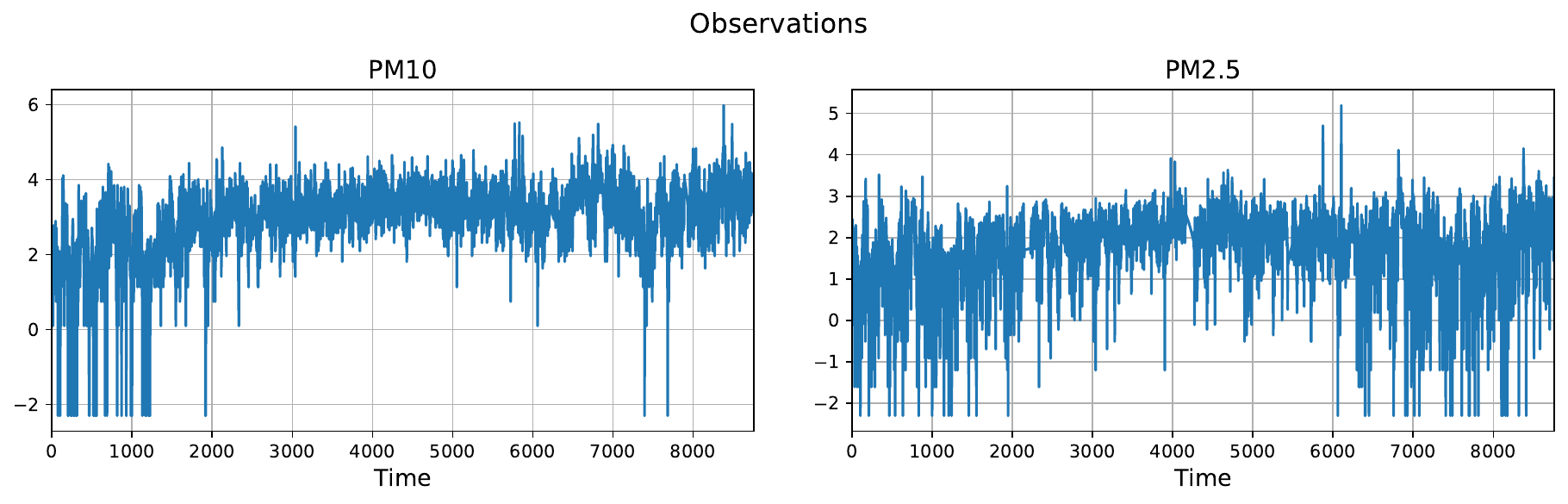}
\caption{Hourly concentrations of $\text{PM}_{10}$ and $\text{PM}_{2.5}$ in Los Angeles over 2017. Both series are transformed using $\log(0.1+\text{PM})$ prior to analysis.}
\label{fig.pm.data}
\end{figure}

We illustrate DC-BATS on a dataset of Los Angeles air quality measurements obtained from the U.S. Environmental Protection Agency (EPA). Aerosol particulates are well known to affect human health, making it important to understand the dynamics of particulate matter (PM) for public health decision-making. The dataset comprises an hourly time series spanning one year, containing two records of different particulates: $\text{PM}_{10}$ (with approximately 1\% missing values) and $\text{PM}_{2.5}$ (with approximately 3.5\% missing values). Missing observations are handled using Kalman smoothing imputation, following \cite{hyndman2008automatic}. After imputation, both PM series are transformed via $\log(0.1+\text{PM})$. The pre-processed data are shown in Figure~\ref{fig.pm.data}. Our goal is to build an interpretable model that captures the dynamics of these time series.

\spacingset{1.1}
\spacingset{\spacing}

Both series exhibit clear heteroskedasticity. To capture the nonstationary variance, and the cross-series correlation, we consider a bivariate GARCH specification with constant conditional correlation \citep{bollerslev1990modelling}, given by
\begin{align} \label{eq.bivariate.garch}
\begin{aligned}
X_t 
& =
\mu + v_t, \quad v_t \sim \N_2 \left (0, H_t \right ),
\\
H_{t,ii} 
& = 
w_i + a_i v^2_{t-1,ii} + b_i H_{t-1,ii},
\\
H_{t,ij} 
& =
r H_{t,ii}^{1/2} H_{t,jj}^{1/2}, \quad i,j = 1, 2,
\end{aligned}
\end{align}
where $X_t \in \RR^2$ denotes the pair ($\text{PM}_{\text{10}}$, $\text{PM}_{\text{2.5}}$) at time $t \in \{1,\dots,T\}$.  This specification assumes a time-independent mean $\mu\in \RR^2$ and a time-dependent innovation $v_t\in\RR^2$ with covariance matrix $H_t$. Each variance component $H_{t,ii}$ follows a univariate GARCH(1,1) process, with intercept $w_i\in \RR_+$, lagged innovation term $v^2_{t-1,ii}$, and lagged variance $H_{t-1,ii}$, with coefficients $a_i,b_i\in \RR_+$. The cross-series correlation is assumed time-independent, captured by $r\in [-1,1]$. 

We specify the diffuse prior distributions $\mathrm{N}(0.5,10^6)$ for $a_i$, $b_i$, $\mu_i$ ($i \in \{1, 2\}$) and $\mathrm{N}(1.0,10^6)$ for $w_i$. As particulate concentrations are known to be positively correlated a priori, we place a $\mathrm{Uniform}(0,1)$ prior on $r$. We draw $10^4$ samples from the posterior $p(a, b, w, \mu \mid X_{1:T})$, where $a = (a_1, a_2)$ and $b=(b_1, b_2)$, obtained using DC-BATS with $k=10$ subsequences and, for comparison, the full-data MCMC. As before, the first half of samples are discarded as burn-in. Sampling from the full posterior required $\sim$24 minutes, whereas DC-BATS required only $\sim$3.8 minutes. Table~\ref{table.pm.CI} reports 95\% credible intervals, showing close agreement between DC-BATS and full-data MCMC.

\spacingset{1.1}
\begin{table}[ht]
\centering
\begin{tabular}{l c c}
\toprule
 & DC-BATS & Full-MCMC \\
\midrule
$a_1$    & $(5.12 \times 10^{-1}, 5.85 \times 10^{-1})$ & $(5.33 \times 10^{-1}, 6.19 \times 10^{-1})$ \\
$a_2$    & $(6.50 \times 10^{-1}, 7.30 \times 10^{-1})$ & $(8.76 \times 10^{-1}, 9.76 \times 10^{-1})$ \\
$b_1$    &$(6.20 \times 10^{-2}, 1.32 \times 10^{-1})$ & $(1.21 \times 10^{-1}, 2.12 \times 10^{-1})$ \\
$b_2$    & $(8.59 \times 10^{-5}, 1.09 \times 10^{-2})$ & $(6.00 \times 10^{-5}, 7.46 \times 10^{-3})$\\
$w_1$    & $(1.22 \times 10^{-1},1.42 \times 10^{-1})$ & $(9.07 \times 10^{-2}, 1.10 \times 10^{-1})$  \\
$w_2$    & $(2.01 \times 10^{-1}, 2.24 \times 10^{-1})$  & $(1.23 \times 10^{-1}, 1.40 \times 10^{-1})$\\
$\mu_1$    & $(3.12,3.14)$ & $(3.25,3.28)$ \\
$\mu_2$    & $(1.95,1.98)$ & $(2.10,2.12)$\\
$r$    & $(2.57 \times 10^{-1},2.78 \times 10^{-1})$ & $(2.32 \times 10^{-1},2.54 \times 10^{-1})$\\ 
\bottomrule
\end{tabular}
\caption{Ninety-five percent credible intervals for the parameters of the bivariate GARCH model \eqref{eq.bivariate.garch} applied to the Los Angeles particulate matter (PM) dataset. Parameter estimation is performed using DC-BATS with $K=10$ subsequences and full-data MCMC.}

\label{table.pm.CI}
\end{table}
\spacingset{\spacing}

\section{Discussion} \label{sec.discussion}

We have proposed a simple divide-and-conquer approach for Bayesian inference with stationary time series. Several natural directions for future work remain. Our theoretical results rely on assumptions of stationarity and fast mixing. It would be interesting to relax these assumptions and develop scalable posterior inference algorithms for nonstationary time series, as well as for series with long-range dependence. While our current algorithm shows promising empirical results in some simulation experiments with nonstationarity, long-range dependence is expected to be more challenging.

Further, we have not considered the problem of optimally parameterizing the subsequence length $m$ and the number of subsequences $K$ for a fixed total sample size $T$. Instead, our experiments focused on challenging regimes where subset sizes are modest and theoretical assumptions are violated (see Supplementary Material for all simulation studies). In practice, for truly massive datasets, MCMC should be run in parallel across subsequences. The optimal choice of $m$ and $K$ will depend on a trade-off between statistical accuracy, computational budget in terms of wall-clock time, number of nodes in a distributed computing network, and the capacity of each node. As a rule of thumb, approximation accuracy should improve with increasing subsequence length, provided the computational budget allows for sufficient MCMC draws per subsequence posterior. Our simulation studies suggest that high accuracy can still be achieved even when subsequences are relatively short.

Two additional directions are especially important: (i) extending the basic divide-and-conquer scheme to allow communication between nodes; and (ii) adapting the algorithm and theory to provide guarantees for fixed, finite subsequence sizes. Both avenues have seen progress outside the time-series setting \citep[e.g.,][]{dai2023bayesian}, and extending them to dependent data remains an open challenge.

\section*{Acknowledgements}

\if1\blind
{Lachlan Astfalck was supported by the ARC ITRH for Transforming energy Infrastructure through Digital Engineering (TIDE; Grant No. IH200100009). Deborshee Sen acknowledges support from SAMSI, (Grant No. DMS-1638521). David Dunson was partially supported by the National Institutes of Health (Grant No. R01ES035625), by the European Research Council under the European Union's Horizon 2020 research and innovation program (Grant No. 856506), and by the Office of Naval Research (Grant No. N00014-21-1-2510).} \fi
\if0\blind
{ \textit{Acknowledgements have been redacted for double blind review.}} \fi

\section*{Code and Data Availability}

Code for all simulation studies is available online at 
\if1\blind
{\texttt{github.com/astfalckl/dcbats.}} \fi
\if0\blind
{ \textit{redacted for double blind review.}} \fi The data used in Section~5 are available at \texttt{epa.gov/outdoor-air-quality-data}.

\subsection*{Declaration of Generative Artificial Intelligence}

This document was reviewed using ChatGPT (models 4o and 5) for minor grammatical suggestions and spelling via VS Code integration. All mathematical content, derivations, and substantive contributions are entirely the authors’ own.

\bibliographystyle{model2-names.bst}
\bibliography{references}

\appendix 
\section{Additional assumptions} \label{sec.additional_assumptions}

Our theoretical results rely on tertiary additional assumptions, and are generalizations to those made in \cite{li2017simple}. In particular, Assumptions \ref{as.domain}, \ref{as.envelop}, \ref{as.fisher} and \ref{as.ui} are generalizations of Assumptions 2, 3, 4, 7 of \cite{li2017simple}, respectively.

\begin{assumption}[Support] \label{as.domain}
For all $t \geq 1$ and all $\theta \in \Theta$, all possible conditional distributions $X_t \mid X_{1:(t-1)}$ have the same support as the stationary distribution of $X_t$.
\end{assumption}

Assumption~\ref{as.domain} ensures that the conditional distributions of $X_t$ given past observations retain the same support as the marginal stationary distribution. This rules out degenerate or absorbing dynamics that would restrict the effective sample space over time. Most classes of time-series models satisfy \cref{as.domain}, including those considered in this paper. 

\begin{assumption}[Envelope] \label{as.envelop}
This assumption consists of three parts.
\begin{enumerate}
\item 
The conditional log-likelihood $\log p_\theta (X_t \mid X_{1:(t-1)})$ is three times differentiable with respect to $\theta$ in a neighbourhood $B_{\delta_0}(\thno) =\{\theta \in \Theta : \|\theta-\thno\|\leq \delno\}$ of $\thno$ for some constant $\delta_0>0$. 
\item 
The first three derivatives of $\log p_\theta (X_t \mid X_{1:(t-1)})$ with respect to $\theta$ are uniformly bounded over $B_{\delta_0}(\thno)$ by a function $M_t(X_{1:t})$, for each $t \geq 1$, such that
\begin{align*}
\sup _{\theta \in B_{\delta_0}(\thno)}\left|\frac{\partial}{\partial \theta_{i}} \log p_\theta (X_t \mid X_{1:(t-1)}) \right|
& \leq 
M_t(X_{1:t}),
\\
\sup_{\theta \in B_{\delta_0}(\thno)}\left| \frac{\partial^2}{\partial \theta_{i} \partial \theta_{j}} \log p_\theta(X_t \mid X_{1:(t-1)}) \right| 
& \leq
M_t(X_{1:t}),
\\
\sup _{\theta \in B_{\delta_0}(\thno)}\left|\frac{\partial^{3}}{\partial \theta_{i} \partial \theta_{j} \partial \theta_{k}} \log p_\theta(X_t \mid X_{1:(t-1)}) \right| 
& \leq 
M_t(X_{1:t}),
\end{align*}
for all indices $i, j, k \in \{1, \dots, d\}$ and all $t \geq 1$.
\item 
The envelope functions satisfy 
\[\limsup\limits_{T \to \infty}\EE_{\thno} \left[T^{-1} \sum_{t=1}^T M_t(X_{1:t})^{4+2\delta}\right] < \infty,\]
where $\delta$ is the same as in \cref{as.mixing}.
\end{enumerate}
\end{assumption}

Together, Assumptions~\ref{as.mixing} and~\ref{as.envelop} jointly impose a trade-off between the mixing rate of the process $\{X_t\}_{t \geq 1}$ and the moment conditions required for the envelope function. Specifically, a larger value of $\delta$ leads to stronger moment conditions in Assumption~\ref{as.envelop}, but relaxes the requirement on the mixing coefficients in Assumption~\ref{as.mixing}, since the summability condition $\sum_{j=1}^\infty \alpha(j)^{\delta/(2+\delta)} < \infty$ becomes easier to satisfy as $\delta$ increases. Assumption~\ref{as.envelop} is straightforward to verify in models where the conditional likelihood depends only on a finite number of past observations. 
\begin{assumption}[Asymptotic local quadratic behavior] \label{as.fisher}
The interchange of order of integration with respect to $\PP_{\thno}$ is valid at $\thno$. The score function $\nabla_\theta \ell_k( \theta)$ is a martingale at $\theta=\thno$ for $m\geq 1$, and 
\begin{equation*}
-T^{-1}\nabla_\theta^2 \ell_T(\thno)
\to 
I(\thno) 
~ \text{in} ~ \PP_{\thno} \text{-probability as } T \to \infty,
\end{equation*}
where $I(\thno)$ is a positive definite matrix.
Finally, for all sufficiently large $m$, the normalized Hessian $-m^{-1} \nabla_\theta^2 \ell_k(\theta)$ is positive definite with eigenvalues bounded below and above by constants for all $\theta\in B_{\delta_0}(\thno)$ and all values of $\Xk$.
\end{assumption}
\cref{as.fisher} generalizes the common local asymptotic quadratic condition to dependent processes. Combined with the moment bounds in \cref{as.envelop}, this ensures that the $L^p$ ergodic theorem \citep[see][]{neumann1932proof} applies to the second derivative of the log-likelihood, and guarantees convergence in $\PP_{\thno}$-probability to the Fisher information matrix $I(\theta_0)$.

\begin{assumption}[Likelihood identifiability] \label{as:likeli}
For any $\delta>0$, there exists an $\epsilon>0$ such that
\begin{equation*}
\lim _{m \to \infty} \PP_{\thno}\left ( \sup _{\theta \in \Theta \, : \, \| \theta - \thno \| \geq \delta}
\frac{\ell_1(\theta)-\ell_1(\thno)}{m}
\leq
-\epsilon \right )
=
1. 
\end{equation*}
\end{assumption}
\cref{as:likeli} is an identifiability condition. This guarantees that $\theta_0$ uniquely maximizes the limiting expected log-likelihood and that subsequence MLEs are consistent. It rules out flat or multimodal likelihood surfaces in the limit, and is a standard prerequisite for local asymptotic quadratic expansions around $\theta_0$.

\begin{assumption}[Prior regularity] \label{as.prior}
The prior density $\pi_0(\theta)$ is continuous at $\thno$; is bounded $0 < \pi_0(\thno) < \infty$; and the second moment of the prior exists: $\int_\theta\|\theta\|^2 \, \pi_0(\theta) \, \dd \theta < \infty$.
\end{assumption}
\cref{as.prior} imposes standard regularity on the prior so that at the true parameter, the posterior is locally dominated by the likelihood. The assumption of finite second moment is required in order to use the $\W_2$ distance to combine the subsequence posteriors.

\begin{assumption}[Uniform integrability] \label{as.ui}
Let $\psi( \bX_{[1]} ) = \EE_{\Pi_m (\dd \theta \mid \bX_{[1]})} \{ K m \| \theta - \thetahat_1\|^2 \}$, where $\EE_{\Pi_m(\dd \theta \mid \bX_{[1]})}$ is the expectation with respect to $\theta$ under the posterior $\Pi_m(\dd \theta \mid \bX_{[1]}) .$ Then there exists an integer $m_0 \geq 1,$ such that $\{\psi(\bX_{[1]}) : m \geq m_0, K \geq 1\}$ is uniformly integrable under $\PP_{\thno}.$ In other words, 
\begin{equation*}
\lim _{C \to \infty} \sup_{m \geq m_0, \, K \geq 1} 
\EE_{{\thno}} \left [ \psi( \bX_{[1]} ) \indicator \{\psi(\bX_{[1]}) \geq C\} \right ]=0,
\end{equation*}
where $\indicator(\cdot)$ is the indicator function.
\end{assumption}
\cref{as.ui} mirrors Assumption 7 in \cite{li2017simple} and is a uniform integrability condition on the posterior second moment of $\theta$ around the subsequence MLE. It requires that the posterior variance, once scaled by the effective sample size $Km$, remains well behaved and does not occasionally take extremely large values, that is, it ensures that posterior variances do not place excessive mass in the tails uniformly over $m$ and $K$.

\end{document}